\documentclass{aastex62}
\usepackage{graphicx}
\usepackage{amsfonts}
\usepackage{tabularx}
\usepackage{mathtools}
\usepackage{bm}
\usepackage{esvect}
\usepackage{natbib}
\usepackage{moreverb}

\submitjournal{RNAAS}
\shorttitle{Generating Density Maps of the Large-scale Structure.}
\shortauthors{Olivia Curtis}

\begin{document}

\title{Fast Generation of Large-Scale Structure Density Maps via Generative Adversarial Networks}
\correspondingauthor{Olivia Curtis}
\email{amcurtis@bu.edu}
\author[0000-0002-0212-4563]{O. Curtis}
\affil{Department of Astronomy \& Institute for Astrophysical Research, Boston University, Boston, MA 02215, USA}
\author{T. G. Brainerd}
\affil{Department of Astronomy \& Institute for Astrophysical Research, Boston University, Boston, MA 02215, USA}

\begin{abstract}
Generative Adversarial Networks (GANs) are a recent advancement in unsupervised machine learning. They are a cat-and-mouse game between two neural networks: [1] a discriminator network which learns to validate whether a sample is real or fake compared to a training set and [2] a generator network which learns to generate data that appear to belong to the training set. Both networks learn from each other until training is complete and the generator network is able to produce samples that are indistinguishable from the training set. We find that GANs are well-suited for fast generation of novel 3D density maps that are indistinguishable from those obtained from N-body simulations. In a matter of seconds, a fully trained GAN can generate thousands of density maps at different epochs in the history of the universe. These GAN-generated maps can then be used to study the evolution of large-scale structure over time.

\end{abstract}

\section{Introduction} \label{sec:intro}

Rather than being distributed randomly throughout the universe, galaxies are found primarily within an interconnected, large-scale network of walls and filaments that stretch for 100's of Mpc (see, e.g., \citealt{1991ARA&A..29..499G} and references therein). 
Between the walls and filaments lie vast, underdense regions known as cosmic voids (see, e.g., \citealt{sanchez2016cosmic}). Voids are the largest structures in the universe, reaching up to $100h^{-1}$~Mpc in diameter, and they have the potential to serve as excellent laboratories for testing the popular $\Lambda$ Cold Dark Matter ($\Lambda$CDM) model of structure formation. Due to their underdense nature, voids are dominated by dark energy and they are only weakly influenced by the non-linear effects of gravity. Because of this, the shapes and distributions of voids provide constraints on the dark energy equation of state, inflationary models, the sum of the neutrino masses, etc. (e.g., \citealt{li2012haloes}; \citealt{clampitt2013voids}; \citealt{sahlen2019cluster}).

Current observational void catalogs (e.g., \citealt{sanchez2016cosmic}; \citealt{mao2017cosmic}) are plagued by small number statistics due to being insufficiently deep for a complete sampling of structure on scales $\gtrsim 1$~Gpc. For example, the Dark Energy Survey (see, e.g., \citealt{dark2005dark}) found only 87 voids in their first 139 sq.\ degree survey (\citealt{sanchez2016cosmic}). The small number of voids in the current catalogs limits our ability to place strong observational constraints on fundamental statistics such as void frequency and mean void radius. Near-future surveys from the Vera C.\ Rubin Observatory, as well as the Nancy Grace Roman and {\it Euclid} satellites, will result in deep maps of the universe that are able to fully sample structure on Gpc-scales. 

Gigaparsec-scale simulations are necessary to determine whether or not $\Lambda$CDM can reproduce the largest structures in the universe. While Gpc-scale N-body mock catalogs do already exist (e.g., \citealt{kim2011new}), the largest voids are so rare that only a few dozen are found in a single Gpc-scale simulation.  Such enormous simulations are computationally expensive to run, making it highly impractical to derive the statistics of the largest voids from these types of catalogs. In order to obtain a complete understanding of Gpc-scale structure in $\Lambda$CDM, a method that can quickly produce catalogs of simulated large-scale structure is required. 

\section{Methods}

\subsection{Generative Adversarial Networks (GANs)}

A Generative Adversarial Network (GAN) (e.g., \citealt{goodfellow2014generative}) offers a way to quickly produce novel catalogs of large-scale structure. GANs are a game between two deep neural networks: [1] a generator, which generates images, and [2] a discriminator, which learns to tell whether an image is original or generated.  The Generator Network takes a random input vector and creates a novel density map that is then judged by the Discriminator Network. If the Discriminator Network deems the sample to be ``fake'', then the Generator learns from its mistakes and tries again. This process continues, with each network learning from the other, until an equilibrium is reached and the generator is able to produce samples that are indistinguishable from the original.
GANs have already found success in astronomy. For example, \cite{rodriguez2018fast} used GANs to produce 2D images of the large-scale structure of $\Lambda$CDM universes, \cite{mustafa2019cosmogan} used GANs to produce novel weak lensing convergence maps, and \cite{schawinski2017generative} used GANs to recover features in Hubble Space Telescope images beyond the deconvolution limit.

Here we present some preliminary results for GAN-generated large-scale structure in $\Lambda$CDM universes. Throughout, we adopt values of the fundamental cosmological parameters that are consistent with a Planck cosmology.

\subsection{Experiments}
We performed two experiments that focused on:
\begin{enumerate}
    \item reproducing the work of \cite{rodriguez2018fast} for the generation of 2D mass distributions, and 
    \item expanding on Experiment~1 to produce 3D distributions of large-scale structure.

\end{enumerate}

For Experiment~1 we used 10 Gadget-2 N-body simulations (see \citealt{springel2005cosmological}) with $512^3$ particles in boxes of sidelength $100h^{-1}$~Mpc. We extracted 15,000 2D images from the simulations in the same fashion described in \cite{rodriguez2018fast} to train a Deep Convolutional GAN (DCGAN). We also utilized the same smoothing function, normalization function, and network architecture that was used by \cite{rodriguez2018fast}.


For Experiment~2 we used a training set consisting of 500 Gadget-2 N-body simulations with $128^3$ particles in boxes of sidelength $100h^{-1}$~Mpc.
We applied a series of random rotations and reflections to the simulations to obtain 15,000 3D density maps to train a DCGAN. Each data cube was voxelated with a Cloud-in-Cell mass assignment scheme, then smoothed and normalized in the same way as the images from Experiment~1. The architecture for this experiment was the same as in Experiment~1, except here 3D convolution layers and half as many filters in the Generator Network's initial layer were used.


\section{Results} \label{sec:results}

\begin{figure}
    \centering
    \includegraphics[width=0.99\textwidth]{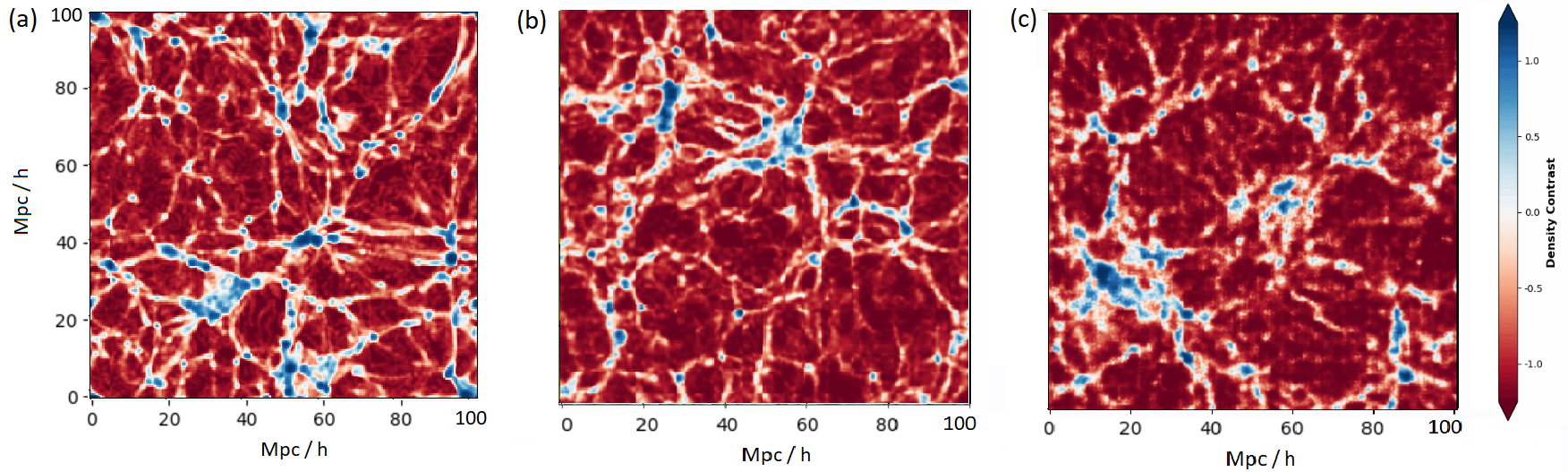}
    \caption{(a) 2D slice from a 3D Gadget-2 N-body simulation. Pixels show the matter density relative to the average density (i.e., the density contrast) and are normalized to the range [-1,1]. (b) Generated 2D image from the DCGAN in Experiment~1. (c) 2D slice from a 3D simulation generated with the DCGAN in Experiment~2.}
    \label{fig:fig_1}
\end{figure}

Our DCGANs were able to reproduce density maps of the large-scale structure. Fig.\ \ref{fig:fig_1} (a) shows an example 2D slice from the N-body simulation training set, Fig.\ \ref{fig:fig_1} (b) shows a generated 2D image of the large-scale structure from Experiment~1 after 21 training epochs, and Fig.\ \ref{fig:fig_1} (c) shows a 2D slice from a generated 3D density map of the large-scale structure from Experiment~2 after 5 training epochs.




\section{Conclusion} \label{sec:conclusion}

GANs are a new advancement in machine learning that pit two neural networks against each other in an adversarial game. The result is a generator network that can produce novel results that are indistinguishable from a training set. \cite{rodriguez2018fast} showed that GANs can produce novel 2D density maps that even experts in the field have trouble distinguishing. The success of GANs in the field of machine vision indicates that GANs could be a useful tool for astronomers, allowing them to quickly generate simulations for the purposes of model testing. Here we showed that we have successfully trained a DCGAN to generate novel 3D density maps of a $\Lambda$CDM universe.  In the future, we will use our DCGAN to produce density maps that are comparable in size to those produced by Gpc-scale N-body simulations.  This will allow us to study the formation and evolution of the largest voids in $\Lambda$CDM universes, and will also allow direct comparisons of future void catalogs to the predictions of $\Lambda$CDM.

\bibliographystyle{aasjournal}
\bibliography{references.bib}


\end{document}